\documentclass[10pt,journal,compsoc,table]{IEEEtran}

\usepackage{todonotes}
\usepackage{xcolor}
\usepackage{xparse}
\usepackage{dblfloatfix}
\usepackage{pgfplots}
\usepackage{tikz}
\usepackage{multirow}
\usepackage{enumitem}
\usepackage[normalem]{ulem}
\usetikzlibrary{matrix, arrows,shapes, calc, fit, positioning, backgrounds}
\usepackage{subcaption}
\pgfplotsset{width=8.25cm,compat=1.9}

\def\BibTeX{{\rm B\kern-.05em{\sc i\kern-.025em b}\kern-.08em
    T\kern-.1667em\lower.7ex\hbox{E}\kern-.125emX}}

\definecolor{customgreen}{rgb}{0.0, 0.4, 0.25}
\newcommand{\sridutt}[1]{\textcolor{black}{#1}}

\newcommand{\linebreakand}{%
  \end{@IEEEauthorhalign}
  \hfill\mbox{}\par
  \mbox{}\hfill\begin{@IEEEauthorhalign}
}

\newcommand*{\scalefactor}{0.5}

\tikzset{
    mymatrix/.style = {matrix of math nodes,
                      nodes in empty cells,
                      nodes={minimum height=2ex, minimum width=1em,
                             inner sep=0pt, outer sep=0pt, anchor=center,
                             draw, very thin,scale=\scalefactor,transform shape},
                      column sep=0pt,
                      row sep=0pt,
                      inner sep=0.5\pgflinewidth, outer sep=1pt,
                      draw, thick,font=\fontsize{6.0}{22.4}\selectfont},
 empty node/.style = {draw,fill=none,scale=\scalefactor},
every label/.append style = {font=\tiny, font=\bfseries, text=black,scale=\scalefactor}
    }
\makeatletter
\def\tikz@lib@matrix@empty@cell{%
\iftikz@lib@matrix@empty%
\node[name=\tikzmatrixname-\the\pgfmatrixcurrentrow-\the\pgfmatrixcurrentcolumn,empty node]{};\fi}
\makeatother

\newcommand{\tableborderwith}{3pt}
\newcommand{\tablescale}{0.24}
\ExplSyntaxOn
\NewDocumentCommand{\he}{m}
 {
  \cellcolor[gray]{ \fp_eval:n { min ( 2*#1, 1 ) } }
  \phantom{#1}
 }
\ExplSyntaxOff
\begin{document}

\title{Understanding the Impact of Input Entropy on FPU, CPU, and GPU Power}

\author{\IEEEauthorblockN{Sridutt Bhalachandra,
Brian Austin, Samuel Williams, and
Nicholas J. Wright}
\\
\IEEEauthorblockA{Lawrence Berkeley National Laboratory, Berkeley, USA\\
\emph{\{sriduttb, baustin, swwilliams, njwright\}@lbl.gov}}}

\makeatletter
\long\def\@IEEEtitleabstractindextextbox#1{\parbox{0.922\textwidth}{#1}}
\makeatother
\IEEEtitleabstractindextext{%
\begin{abstract}
Power is increasingly becoming a limiting resource in high-performance, GPU-accelerated computing systems. 
Understanding the range and sources of power variation is essential in setting realistic bounds on rack and system peak power, and developing techniques that minimize energy. 
While variations arising during manufacturing and other factors like algorithm among others have been previously studied, this work shows that the program inputs can also severely impact the power consumed not only on the GPU but also CPUs.
Power variations of up to 67\% were observed on an NVIDIA Ampere A100 GPU for the same algorithm (DGEMM benchmark) and input size with different matrix values.
Our investigation shows that the values used as matrix elements, their position, and their uniqueness strongly influence power consumption.
The implications of this result on supercomputer performance and energy efficiency are further discussed.
% The abstract goes here.
\end{abstract}
}

\maketitle

\IEEEdisplaynontitleabstractindextext

\IEEEpeerreviewmaketitle

\ifCLASSOPTIONcompsoc
\IEEEraisesectionheading{\section{Introduction}\label{sec:introduction}}
\else
\section{Introduction}
\label{sec:introduction}
\fi
The first exascale system, Frontier~\cite{frontier}, has achieved an impressive energy efficiency of 52 GFLOPS/W (19 pJ/FLOP) meeting the exascale computing study goals~\cite{kogge2008exascale}.
While this indicates a marked improvement in the energy efficiency of hardware architectures, it may not necessarily be adequate.
Future system peak performance will be predominantly limited by how efficiently available power budgets are managed.
In particular, large power swings in supercomputers can detract from achieving sustained performance~\cite{bhalachandra2021understanding}.
Consequently, understanding the sources of power variation and inter-process correlations in power will be vital to designing future systems that are both high-performance and energy-efficient.

Variations in the power consumption of systems have been a well-known and widely studied topic of HPC research.
Manufacturing variations that arise during the fabrication process can induce variations in the performance and power of otherwise identical processors~\cite{dblp:conf/ipps/rountreeasls12}.
Component placement too can cause deviations due to varying thermal efficiencies\cite{marathe2017towards}.
However, the cause of these differences is not limited to hardware.
Software variations emanating from varying data locality, compiler optimizations, and context switching, have been shown to cause 20\% variations in power in general and over a 2$\times$ variation in power in extreme case~\cite{porterfield2013power}.
The presence of this variability can hinder performance and energy-tuning efforts leading to a choice of sub-optimal versions of code~\cite{porterfield2016variability}.

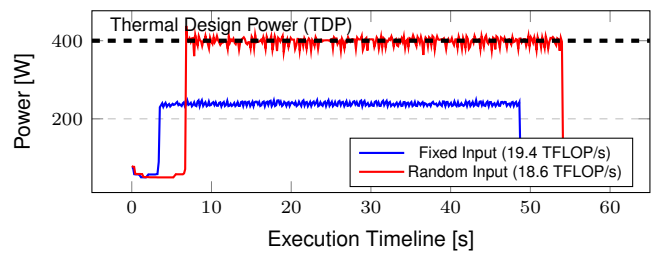
\begin{figure}[tb]
\centering
\begin{tikzpicture}
\begin{axis}[
    width=9cm,
    height=4cm,
    xlabel={Execution Timeline [s]},
    ylabel={Power [W]},
    xmin=-5, xmax=65,
    legend pos=south east,
    ymajorgrids=true,
    grid style=dashed,
    legend style={nodes={scale=\scalefactor, transform shape,font=\large}}, 
    label style={font=\footnotesize},
    tick label style={font=\scriptsize}
]

\addplot[
    color=blue,
    mark=none,
    thick
    ]
    coordinates {
    (0.0, 75.46) (0.155, 75.46) (0.303, 58.27) (0.463, 57.74) (0.602, 57.74) (0.76, 57.74) (0.92, 57.5) (1.16, 50.29) (1.326, 50.29) (1.616, 50.05) (1.782, 50.05) (2.091, 57.74) (2.244, 57.5) (2.393, 57.5) (2.534, 57.5) (2.673, 57.5) (2.808, 57.5) (2.952, 57.5) (3.087, 58.27) (3.222, 58.51) (3.382, 89.83) (3.52, 229.71) (3.657, 234.33) (3.82, 239.95) (3.956, 230.76) (4.094, 236.9) (4.233, 241.0) (4.399, 231.23) (4.545, 235.33) (4.681, 237.9) (4.821, 231.52) (4.984, 235.33) (5.123, 238.42) (5.26, 235.61) (5.417, 236.9) (5.555, 240.47) (5.696, 243.28) (5.844, 235.33) (5.981, 238.19) (6.119, 235.33) (6.255, 232.81) (6.389, 239.43) (6.533, 242.29) (6.67, 231.76) (6.808, 235.61) (6.967, 246.56) (7.106, 235.61) (7.24, 237.37) (7.404, 241.0) (7.542, 232.04) (7.68, 235.61) (7.829, 242.29) (7.965, 234.09) (8.104, 232.51) (8.242, 243.81) (8.378, 247.15) (8.519, 232.51) (8.656, 242.0) (8.794, 247.62) (8.955, 233.04) (9.095, 238.19) (9.235, 242.52) (9.398, 232.28) (9.535, 234.57) (9.67, 244.05) (9.808, 234.33) (9.948, 235.61) (10.09, 240.71) (10.229, 242.76) (10.369, 235.33) (10.526, 242.29) (10.671, 233.56) (10.822, 233.56) (10.962, 241.0) (11.099, 242.29) (11.24, 232.04) (11.377, 239.19) (11.542, 243.05) (11.679, 232.81) (11.842, 236.14) (11.979, 245.86) (12.122, 232.51) (12.262, 233.56) (12.401, 238.67) (12.551, 233.04) (12.691, 235.61) (12.854, 240.71) (12.994, 232.51) (13.131, 235.33) (13.283, 242.52) (13.42, 233.04) (13.56, 233.04) (13.699, 242.76) (13.836, 244.05) (13.981, 233.04) (14.12, 243.53) (14.254, 243.76) (14.414, 233.56) (14.552, 242.76) (14.693, 244.05) (14.858, 232.51) (14.997, 238.19) (15.137, 243.81) (15.273, 232.81) (15.409, 242.52) (15.547, 244.05) (15.686, 243.76) (15.825, 237.14) (15.98, 243.05) (16.116, 232.81) (16.253, 234.57) (16.412, 242.52) (16.553, 244.33) (16.693, 232.28) (16.855, 242.29) (16.992, 243.81) (17.129, 235.61) (17.278, 245.1) (17.414, 246.38) (17.553, 232.51) (17.691, 237.9) (17.829, 247.15) (17.975, 232.81) (18.111, 234.85) (18.249, 246.86) (18.41, 247.38) (18.549, 238.42) (18.687, 244.05) (18.851, 232.81) (18.989, 235.85) (19.124, 243.53) (19.26, 240.71) (19.397, 233.04) (19.535, 241.24) (19.671, 243.05) (19.805, 232.81) (19.952, 239.95) (20.096, 242.76) (20.231, 239.95) (20.398, 236.38) (20.533, 242.52) (20.674, 232.51) (20.841, 235.61) (20.979, 244.05) (21.118, 238.19) (21.26, 232.81) (21.401, 244.57) (21.537, 247.38) (21.676, 234.33) (21.813, 241.76) (21.95, 243.05) (22.09, 232.28) (22.233, 237.9) (22.374, 243.28) (22.515, 232.81) (22.676, 241.48) (22.821, 244.33) (22.96, 236.14) (23.114, 241.0) (23.252, 247.62) (23.39, 232.28) (23.527, 234.33) (23.665, 242.29) (23.804, 236.14) (23.941, 233.28) (24.077, 241.48) (24.234, 244.05) (24.372, 232.51) (24.51, 240.47) (24.674, 247.38) (24.81, 233.56) (24.947, 238.67) (25.094, 243.81) (25.229, 232.28) (25.365, 235.85) (25.503, 243.05) (25.642, 234.57) (25.783, 233.04) (25.919, 243.81) (26.058, 245.86) (26.218, 235.09) (26.355, 242.29) (26.493, 243.53) (26.657, 232.51) (26.795, 238.67) (26.931, 244.81) (27.069, 233.04) (27.207, 233.8) (27.35, 242.29) (27.485, 233.56) (27.624, 237.37) (27.771, 243.81) (27.912, 245.1) (28.051, 236.9) (28.217, 243.05) (28.355, 234.09) (28.492, 233.28) (28.644, 244.81) (28.779, 246.86) (28.919, 232.51) (29.058, 242.29) (29.196, 245.86) (29.335, 234.09) (29.471, 237.37) (29.608, 240.47) (29.767, 232.28) (29.906, 233.8) (30.042, 241.48) (30.209, 233.56) (30.347, 233.04) (30.482, 243.28) (30.647, 245.33) (30.783, 235.61) (30.943, 242.52) (31.083, 233.04) (31.222, 233.04) (31.387, 243.28) (31.527, 246.38) (31.665, 235.09) (31.83, 242.0) (31.968, 246.1) (32.103, 232.81) (32.241, 241.0) (32.379, 243.28) (32.525, 232.81) (32.664, 236.14) (32.803, 238.42) (32.958, 233.04) (33.095, 237.9) (33.232, 241.48) (33.394, 232.51) (33.531, 238.19) (33.667, 242.76) (33.819, 232.81) (33.957, 233.28) (34.092, 242.76) (34.23, 243.05) (34.368, 232.51) (34.51, 240.24) (34.647, 241.48) (34.781, 233.04) (34.937, 239.95) (35.078, 243.05) (35.214, 232.51) (35.377, 235.61) (35.512, 243.05) (35.649, 233.28) (35.797, 232.81) (35.933, 242.76) (36.071, 239.95) (36.213, 235.09) (36.35, 239.95) (36.486, 242.76) (36.623, 233.56) (36.76, 239.95) (36.92, 238.19) (37.059, 232.81) (37.193, 236.9) (37.355, 238.42) (37.493, 232.81) (37.631, 240.71) (37.788, 243.53) (37.922, 237.66) (38.058, 237.14) (38.197, 238.95) (38.336, 232.81) (38.471, 235.33) (38.609, 242.0) (38.746, 236.38) (38.903, 233.04) (39.039, 244.57) (39.176, 244.33) (39.34, 232.81) (39.477, 243.05) (39.615, 243.28) (39.766, 232.81) (39.902, 237.14) (40.042, 241.0) (40.179, 232.81) (40.329, 234.57) (40.476, 244.33) (40.612, 235.85) (40.751, 232.51) (40.918, 242.76) (41.059, 232.81) (41.198, 234.33) (41.348, 239.95) (41.486, 234.85) (41.624, 233.04) (41.762, 236.38) (41.896, 244.33) (42.058, 232.81) (42.196, 242.24) (42.335, 244.33) (42.481, 232.81) (42.625, 237.14) (42.763, 243.05) (42.905, 232.81) (43.044, 235.61) (43.195, 247.62) (43.33, 234.33) (43.471, 232.81) (43.636, 243.81) (43.776, 244.57) (43.918, 234.85) (44.063, 242.76) (44.201, 234.09) (44.344, 233.04) (44.484, 241.0) (44.625, 244.33) (44.777, 232.51) (44.915, 242.76) (45.055, 242.99) (45.22, 233.04) (45.357, 237.37) (45.496, 239.95) (45.645, 232.51) (45.785, 237.14) (45.924, 243.28) (46.064, 233.56) (46.2, 233.28) (46.35, 243.05) (46.488, 243.28) (46.625, 239.19) (46.77, 243.28) (46.907, 243.05) (47.044, 233.28) (47.21, 237.9) (47.349, 238.95) (47.487, 233.04) (47.649, 242.0) (47.786, 243.28) (47.924, 233.04) (48.071, 240.47) (48.209, 243.28) (48.347, 233.04) (48.484, 235.61) (48.623, 242.29) (48.768, 84.96) (48.908, 78.8) (49.046, 77.51) (49.213, 76.22) (49.603, 75.99) (49.985, 75.7) (50.22, 75.7) (50.364, 75.99)
    };
    \addlegendentry{Fixed Input (19.4 TFLOP/s)}

\addplot[
    color=red,
    mark=none,
    thick
    ]
    coordinates {
    (0.0, 79.04) (0.163, 77.23) (0.297, 76.99) (0.434, 59.27) (0.596, 58.74) (0.736, 58.74) (0.873, 58.74) (1.034, 58.74) (1.169, 58.51) (1.306, 58.51) (1.453, 51.34) (1.593, 51.06) (1.732, 50.82) (1.876, 50.82) (2.016, 50.82) (2.179, 50.53) (2.321, 50.29) (2.461, 50.29) (2.623, 50.29) (2.764, 50.29) (2.906, 50.29) (3.072, 50.29) (3.213, 50.05) (3.353, 50.05) (3.51, 50.05) (3.65, 50.05) (3.791, 50.05) (3.928, 50.05) (4.067, 50.05) (4.237, 50.05) (4.415, 50.05) (4.65, 50.05) (4.823, 50.05) (5.046, 50.05) (5.214, 50.05) (5.532, 58.27) (5.693, 57.98) (5.843, 57.98) (5.98, 57.98) (6.136, 57.98) (6.272, 57.98) (6.408, 57.98) (6.56, 60.32) (6.699, 71.07) (6.834, 437.93) (6.973, 402.66) (7.11, 398.81) (7.254, 405.76) (7.394, 397.05) (7.531, 405.99) (7.686, 406.99) (7.826, 360.5) (7.964, 414.17) (8.128, 397.58) (8.266, 387.31) (8.403, 411.63) (8.555, 401.66) (8.692, 398.34) (8.829, 386.55) (8.964, 398.81) (9.102, 397.29) (9.245, 412.65) (9.385, 400.38) (9.521, 393.73) (9.687, 380.18) (9.823, 408.04) (9.962, 399.1) (10.128, 392.2) (10.267, 408.79) (10.404, 396.01) (10.544, 401.38) (10.684, 411.89) (10.824, 401.14) (10.963, 395.77) (11.102, 389.4) (11.262, 394.72) (11.453, 395.53) (11.599, 399.1) (11.738, 400.62) (11.882, 396.53) (12.036, 379.13) (12.173, 411.89) (12.31, 394.72) (12.474, 402.9) (12.611, 399.86) (12.751, 405.23) (12.907, 394.25) (13.043, 390.92) (13.18, 400.38) (13.321, 400.1) (13.46, 377.3) (13.603, 395.53) (13.744, 401.66) (13.885, 405.23) (14.053, 408.79) (14.19, 404.18) (14.329, 393.73) (14.486, 390.64) (14.624, 396.3) (14.771, 396.3) (14.906, 380.94) (15.042, 410.32) (15.202, 403.42) (15.341, 381.18) (15.479, 415.17) (15.641, 396.01) (15.777, 397.05) (15.915, 397.29) (16.082, 399.62) (16.219, 402.66) (16.36, 395.53) (16.505, 409.04) (16.645, 398.34) (16.786, 379.87) (16.928, 410.32) (17.068, 400.86) (17.212, 400.62) (17.351, 401.91) (17.492, 411.89) (17.652, 376.61) (17.795, 392.97) (17.933, 411.6) (18.084, 400.86) (18.221, 373.53) (18.363, 407.51) (18.501, 403.95) (18.637, 397.81) (18.787, 405.47) (18.925, 412.12) (19.062, 370.72) (19.229, 393.2) (19.366, 396.01) (19.504, 404.18) (19.662, 384.03) (19.798, 395.53) (19.935, 407.27) (20.071, 403.19) (20.211, 402.66) (20.351, 402.66) (20.487, 378.66) (20.628, 387.83) (20.79, 404.71) (20.926, 404.47) (21.064, 375.02) (21.228, 403.42) (21.368, 402.9) (21.506, 386.84) (21.651, 406.51) (21.791, 400.1) (21.932, 379.87) (22.067, 399.86) (22.203, 405.76) (22.353, 400.86) (22.491, 375.33) (22.633, 402.43) (22.791, 397.58) (22.929, 387.31) (23.067, 409.04) (23.229, 401.14) (23.371, 379.42) (23.51, 398.05) (23.651, 402.43) (23.785, 402.9) (23.923, 398.58) (24.061, 400.86) (24.2, 400.86) (24.351, 401.14) (24.489, 401.14) (24.627, 401.14) (24.791, 399.86) (24.933, 402.43) (25.095, 404.18) (25.235, 380.63) (25.373, 408.56) (25.538, 404.18) (25.676, 405.47) (25.812, 391.45) (25.95, 397.29) (26.091, 400.38) (26.233, 375.49) (26.374, 405.47) (26.514, 399.1) (26.675, 380.39) (26.817, 398.58) (26.96, 396.01) (27.124, 409.32) (27.267, 402.14) (27.408, 397.81) (27.551, 403.42) (27.69, 404.18) (27.837, 403.19) (27.978, 403.19) (28.119, 379.58) (28.272, 408.04) (28.407, 403.71) (28.544, 403.19) (28.713, 403.42) (28.85, 411.89) (28.989, 376.3) (29.148, 386.07) (29.289, 404.18) (29.427, 407.27) (29.562, 374.73) (29.701, 402.14) (29.843, 403.71) (29.982, 403.42) (30.124, 403.71) (30.289, 403.19) (30.428, 384.2) (30.571, 394.2) (30.724, 408.56) (30.86, 403.42) (30.995, 379.87) (31.134, 395.77) (31.273, 410.6) (31.419, 383.2) (31.56, 407.51) (31.699, 404.47) (31.85, 382.74) (31.986, 399.81) (32.127, 398.34) (32.293, 405.23) (32.432, 402.43) (32.572, 410.84) (32.723, 400.38) (32.858, 385.03) (32.998, 400.86) (33.137, 403.95) (33.274, 388.05) (33.416, 386.01) (33.554, 401.14) (33.689, 395.77) (33.848, 405.99) (33.991, 396.77) (34.13, 410.84) (34.291, 381.39) (34.431, 399.33) (34.568, 402.9) (34.704, 384.2) (34.842, 406.51) (35.009, 404.47) (35.146, 403.19) (35.289, 413.93) (35.43, 406.23) (35.572, 393.97) (35.71, 386.07) (35.851, 409.56) (35.989, 403.71) (36.134, 394.25) (36.277, 409.56) (36.416, 398.81) (36.575, 363.5) (36.713, 397.05) (36.851, 408.27) (37.015, 403.71) (37.155, 385.27) (37.296, 409.04) (37.44, 405.76) (37.577, 388.59) (37.722, 412.12) (37.859, 408.79) (38.006, 377.61) (38.159, 400.86) (38.299, 411.08) (38.437, 403.19) (38.598, 402.14) (38.736, 403.19) (38.873, 403.42) (39.021, 403.71) (39.158, 406.23) (39.296, 395.25) (39.435, 378.66) (39.571, 398.58) (39.713, 404.47) (39.853, 396.77) (39.99, 397.58) (40.152, 406.51) (40.288, 395.25) (40.428, 388.59) (40.594, 403.42) (40.733, 403.42) (40.87, 382.68) (41.009, 410.32) (41.146, 400.62) (41.284, 397.81) (41.421, 379.66) (41.56, 407.27) (41.718, 396.77) (41.856, 381.39) (41.993, 411.36) (42.157, 404.71) (42.292, 379.13) (42.431, 383.96) (42.585, 399.62) (42.724, 402.43) (42.866, 394.72) (43.003, 412.12) (43.142, 403.71) (43.285, 398.81) (43.427, 402.43) (43.566, 403.95) (43.734, 378.89) (43.882, 411.89) (44.02, 399.33) (44.165, 385.55) (44.304, 407.51) (44.441, 413.12) (44.59, 396.3) (44.727, 392.2) (44.886, 403.42) (45.023, 399.33) (45.16, 375.05) (45.324, 408.04) (45.462, 405.47) (45.601, 377.9) (45.747, 411.36) (45.885, 408.56) (46.03, 397.29) (46.173, 381.46) (46.318, 408.56) (46.483, 405.23) (46.623, 399.62) (46.76, 407.51) (46.917, 397.05) (47.053, 382.46) (47.189, 402.66) (47.334, 403.95) (47.471, 396.77) (47.61, 390.4) (47.752, 408.79) (47.891, 396.3) (48.044, 397.81) (48.184, 404.18) (48.321, 402.9) (48.487, 379.13) (48.626, 405.23) (48.767, 405.76) (48.913, 396.01) (49.056, 401.91) (49.194, 404.18) (49.331, 403.95) (49.474, 386.84) (49.626, 397.81) (49.762, 412.65) (49.902, 385.55) (50.067, 409.04) (50.206, 403.95) (50.349, 395.01) (50.494, 408.27) (50.632, 395.25) (50.775, 375.86) (50.917, 408.79) (51.056, 397.81) (51.212, 399.86) (51.347, 382.74) (51.491, 408.56) (51.65, 395.53) (51.795, 381.46) (51.933, 411.36) (52.096, 398.05) (52.235, 379.42) (52.377, 396.53) (52.525, 406.99) (52.662, 405.99) (52.818, 409.56) (52.957, 405.99) (53.092, 394.25) (53.245, 388.36) (53.38, 406.23) (53.517, 403.95) (53.681, 374.28) (53.822, 407.27) (53.96, 397.58) (54.114, 87.79) (54.248, 84.68) (54.386, 82.15) (54.529, 78.52) (54.904, 78.04) (55.285, 77.51) (55.518, 77.51) (55.655, 77.51) 
    };
    \addlegendentry{Random Input (18.6 TFLOP/s)}

\addplot[black,domain = -5:65,dashed,ultra thick] {400} node[above,pos=0.25,scale=\scalefactor] {\Large Thermal Design Power (TDP)};  
  
\end{axis}
\end{tikzpicture}
\caption{Timeline showing the power consumption for 100 DGEMMs with fixed and random inputs on an NVIDIA A100 GPU (Input Size: 16384$\times$16384)}
\label{fig:DGEMM_FixedVsRandom_Timeline_Power}
\end{figure}

Benchmarking power variability can play a critical role in system sizing and rack density.  
Benchmarks that underestimate peak GPU power can result in nodes, racks, and systems that draw substantially higher peak power than expected. 
Figure~\ref{fig:DGEMM_FixedVsRandom_Timeline_Power} shows NVIDIA A100 Ampere GPU power over time when running dense matrix-matrix multiplications (DGEMM) for different input matrices of dimension 16K.
The \emph{fixed input} case uses the same value for all elements of each input matrix, while in the \emph{random} case each element of each matrix is a randomly generated number.
The average power consumed by the \emph{random} case ($\approx$398 W) is about 67\% higher than the \emph{fixed} case ($\approx$239 W).
In addition to a higher power, the performance of the \emph{random} case is nearly 4\% lower (18.6 TFLOPS vs. 19.4 TFLOPS).
These observations suggest that, in addition to input size, the entropy of matrix values strongly influences the power as well as the performance of a GPU.  

% \begin{figure}[ht]
%     \centering
%     \includegraphics[width=0.95\linewidth]{figures/DGEMM_FixedVsRandom_Timeline_Power.png}
%   \caption{
%   Timeline showing the power consumption of DGEMM with fixed and random inputs on Nvidia A100 GPU (Input Size: 16384 x 16384) \textcolor{red}{use PM GPU data, fix labels}
%   }
%   \label{fig:DGEMM_FixedVsRandom_Timeline_Power}
% \end{figure}

\begin{figure}[tb]
\centering
\includegraphics[width=0.8\linewidth,height=2.9cm]{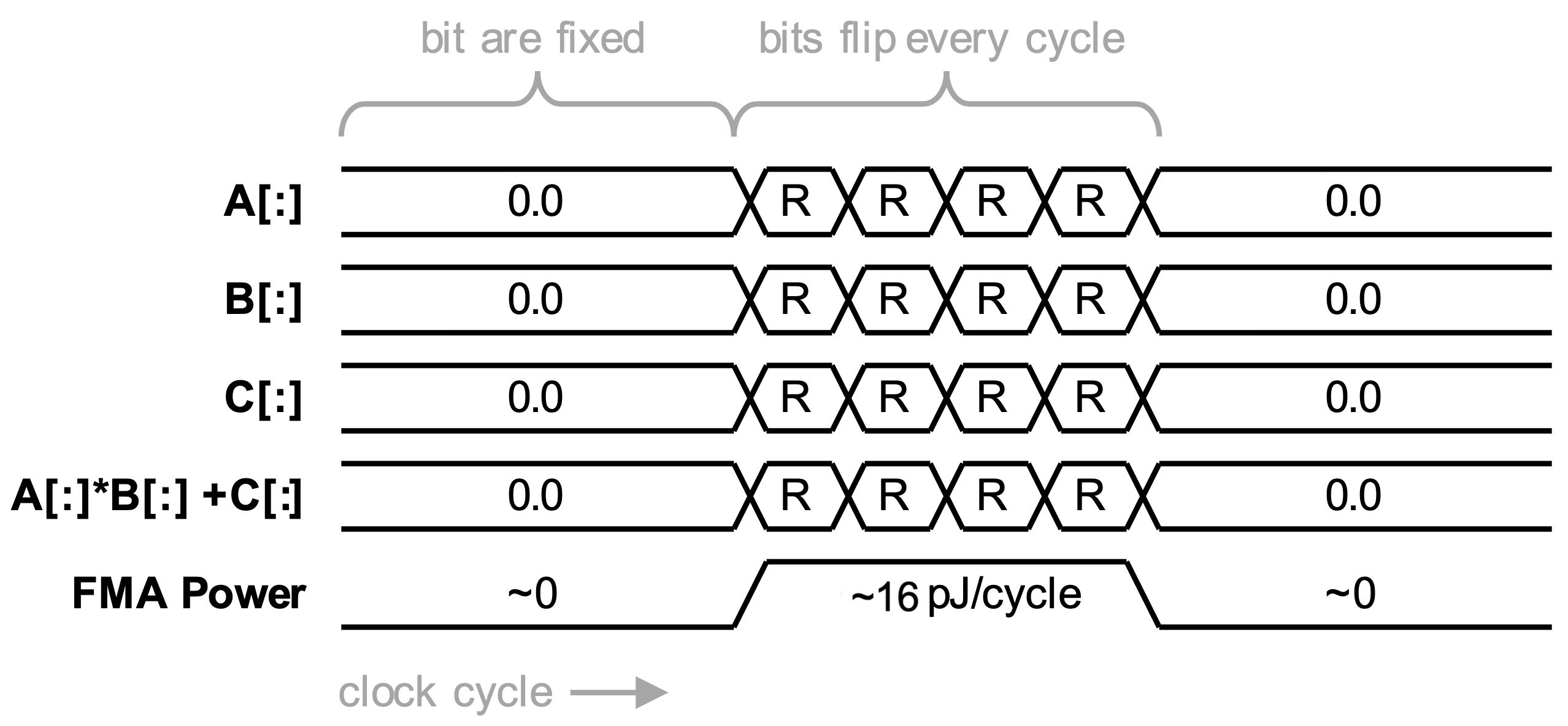}
\caption{Stylized cycle-by-cycle timeline of FMA inputs, output, and power.  FMA power is only high when inputs change.}
\label{fig:cyclebycycle}
\end{figure}

CMOS circuit power is dominated by two terms:  static power and dynamic power.  Whereas static power is a constant drain, dynamic power (or switching power) arises from transistors turning off and on and wire/gate capacitance being charged and discharged.
Datapath functional units (e.g. multipliers, register files, load-store units, etc...) are built from CMOS logic gates.
Changes to functional unit inputs produce dynamic power from each logic gate whose inputs change.
Figure~\ref{fig:cyclebycycle} visualizes how any time an input value changes from one cycle to the next, dynamic energy is consumed. 
We can thus theorize that the 159~W difference between the red (random) and blue (fixed) curves in Figure~\ref{fig:DGEMM_FixedVsRandom_Timeline_Power} is due to the dynamic datapath switching energy.  We estimate (159~W / 19.4~TFLOP/s) an upper bound on FPU energy of about 8pJ/FLOP.  Despite the A100 GPU being built on a 7~nm process, this estimate is remarkably close to the 2018 estimates of what an 11~nm process might provide~\cite{shalf2020future}.  Our estimate also includes amortized DRAM, cache, and register file energy.

We explore the relationship between power and FMA input values in the context of DGEMM on GPUs and CPUs.  We attain insights into the value of tensor cores on FMA energy as well as implications on future GPU and system architectures. All experimental data is available on Zenodo with open access (https://doi.org/10.5281/zenodo.7081498).

\section{Experimental Setup}
\label{sec:setup}
\input{text/figures/Figure_Tabular_BlockPattern_independentAndfixed_combined}
\input{text/figures/Figure_Tabular_SparsePattern_independentAndfixed_combined}
The hardware platform for this study is the Perlmutter supercomputer located at NERSC.
Perlmutter debuted as \#5 on the Top500 and \#6 on the Green500 lists in June 2021. 
It was also the most energy-efficient of the top 10 entries in the Top500 list.
Perlmutter is an HPE EX system comprised of 4,608 compute nodes with a peak performance of around 71~\sridutt{PetaFLOPs}, augmented by a 35~PB all-flash storage system,
all connected by an HPE Cray Slingshot interconnect.
Perlmutter's primary compute partition has 1,536 GPU-accelerated nodes with four NVIDIA A100 Ampere GPUs (160~GB of HBM2 memory in total) and one AMD EPYC 7763 (Milan) processor (512 GB of DDR4 memory).  Each GPU includes 3,456 FP64 FPUs capable of one fused multiply-add (FMA) per cycle.
It also has a secondary partition of 3,072 CPU-only nodes
each with two AMD EPYC 7763 (Milan) processors and 512 GB of DDR4 memory in total.
The Thermal Design Powers (TDPs) for the GPU and CPU are 400~W and 280~W respectively.

The highest resolution power measurements for Perlmutter can be obtained from NVIDIA's System Management Interface (nvidia-smi) or Data Center GPU Manager (DCGM) on each GPU with an interval of one millisecond.
We use nvidia-smi in our experiments with a sampling interval of 100 milliseconds.
CPU-level power measurements are obtained through Cray's power monitoring~(PM) interface~\cite{martin2014cray}.

\section{Methodology \& Evaluation}
\label{sec:evaluation}
DGEMM is the BLAS function that performs
double-precision dense matrix-matrix multiplication.
Optimized implementations of DGEMM
are provided by most processor vendors
and %typically 
approach a processor's peak floating-point performance.
DGEMM calculates the product of double-precision matrices as
\[C\Leftarrow \alpha A*B + \beta C\]

The GPU version of DGEMM from the NERSC Proxies suite~\cite{nersc-proxies} is configured to use  the \texttt{DGEMM} implementation in the NVIDIA cuBLAS Library (v 11.7). %and is used in all our GPU experiments. 
We hypothesize that the higher power consumed by DGEMM in Section~\ref{sec:introduction} with random inputs is caused by a higher number of bit flips.
Each of the 192 input bits to each of the 3456 FP64 FMAs has a chance of flipping from one cycle to the next.  For a multiplier, unlike a bitwise operation, flipping some bits will cause the FMA to burn more energy than flipping others.
Each of those FMAs is time multiplexed among multiple 4 threads in a CUDA warp, time multiplexed across multiple warps in a thread block, and time multiplexed across multiple warps running concurrently on each SM.

\subsection{Matrix Pattern Generators}
Three levels of time multiplexing make it very difficult to analytically construct a matrix whose elements, when mapped to threads and thread blocks, will result in a particular switching frequency across all the FP64 FMAs on an A100 GPU.
As such, we modified the NERSC benchmark to accept different input patterns that we generate with the aim of producing matrices that strongly differentiate dynamic power through the GPU compute and memory subsystems.

The input patterns used in our evaluation can be grouped into two categories - \emph{block} and \emph{sparse}.
In a \emph{block} pattern for populating the DGEMM input matrices, the random values are arranged into 1 to N blocks ($2^n, 0 \leq n \leq log_2 N$) as shown in Figures~\ref{fig:blockPattern_RowCol_Diagram} as block rows and columns, and in Figure~\ref{fig:blockPattern_Diagonal_Diagram} in a checkerboard pattern.
Aside from the fully random base case, all generated matrices are 50\% random and 50\% zero.  Examination of the dot products of rows of $A$ and columns of $B$ highlight patterns of random-random, random-zero, and zero-zero multiplications.  Generalization to high-performance tiled DGEMM implementations produces small random-random, random-zero, and zero-zero matrix-matrix multiplications.  

In a \emph{sparse} pattern for populating the input matrices, the number of random values is increased from $(N^2)/N$ to $(N^2)$ elements ($(N*N)/2^{(log_2 N -n)}, 0 \leq n \leq log_2 N$) as shown in Figures~\ref{fig:sparsePattern_RowCol_Diagram}~\&~\ref{fig:sparsePattern_Diagonal_Diagram} along rows and columns or blocked diagonals (i.e. checkerboard).
The resultant dot products or tiled matrix multiplications are initially multiplications by mostly zeros (thus propagating zeros in the dot product accumulations and also zeroing adder energy) but tend towards fully random.

Whereas random-random multiplications produce substantial FPU energy, we can define four additional cases by taking the aforementioned \emph{block} and \emph{sparse} random patterns by replacing the random values with a fixed (across all matrix entries) random number.  Doing so will minimize memory and multiplier energy as values (bits) do not change.  However, accumulation (adder) energy will persist.
The random values for input matrices (A and B) are generated as \textit{rand() / RAND\_MAX} and have a range between 0 and 1.
The $\alpha$ and $\beta$ values are set to 1.0 with the output matrix C initialized to 0.0.
In the \emph{fixed input} case, the values for A, B, C, $\alpha$, and $\beta$ are 2.0, 0.5, 1.0, 1.0, and 1.0 respectively.

Our benchmark performs 100 DGEMMs ensuring each configuration has an execution time of about a minute.
Both GPU and CPU experiments were run on three different Perlmutter nodes and each experiment was repeated three times.
The variation in power across the nodes was less than 2\%, and thus the mean values (of three runs) from only one of the nodes are reported.
To avoid any startup effects, each experiment was also preceded by a warm-up run lasting about a minute.

\subsection{GPU Results}
Figure~\ref{fig:blockPattern_RowCol_result} shows the power consumption for the \emph{block} row/column pattern with independent and fixed random values (input size of 16K).
It can be seen that the power for 50\% random matrices (red) increases as the random values become more evenly distributed --- a pattern that is intended to produce high cycle-to-cycle bit entropy.
The 50\% random matrices that are the most evenly distributed (at $2^{14}$ blocks) are shown to consume as much power as a fully random matrix.  
Conversely, if one were to replace the random nonzero matrix entries with a common nonzero value (blue), then power drops precipitously.
As warps march through rows or columns, they read and execute multiplications with the same values over and over.
However, as one approaches matrices with high entropy, values change from row to row of $A$ or column to column of $B$ producing increased power.
Nevertheless, the fixed value row/column power reaches only about 75\% of TDP.
Ultimately, these figures highlight that FPU energy is tied less to the values of a matrix and more to how those values change from one column/row to the next.

Figure~\ref{fig:blockPattern_Diagonal_result} presents power for diagonally-structured matrices.  Compared to a fully random matrix, the N/2$\times$N/2 block structure performs many multiplications by zero submatrices, but still incurs power thru the memory/register/FPU subsystem as random values are propagated without any gating/holding when a zero is observed.  
The fixed value diagonal pattern ultimately consumes near TDP power as matrix values are very different for adjacent rows or columns.
Both Figures~\ref{fig:blockPattern_RowCol_result} and \ref{fig:blockPattern_Diagonal_result} have equal 50\% sparsity (aside from the first data point) and generally observe similar trends of increasing power where values change faster than every 16 rows or columns.

Figures~\ref{fig:sparsePattern_RowCol_result} and \ref{fig:sparsePattern_Diagonal_result} show that TDP increases as sparsity (the number of zero rows/columns/diagonals) decreases.
Similar to the \emph{block} patterns,  power consumption increases quickly once more than about 6\% of the matrix is random.
For the diagonal(checkerboard pattern), TDP is attained when the matrix reaches 50\% sparsity as there is maximal variation between adjacent rows and columns.
When the random values of the matrices are replaced with a constant, power drops precipitously for the blocked row/column structure, but not for the diagonal format.
For any vectorized implementation, the constant checkerboard format is effectively the same as the random checkerboard format (striding in rows or columns produces high variability).
Once again, as entropy increases, power increases.  However, it is clear that the actual pattern will define how much power can vary and how the presence of constant values will further reduce power.

\subsection{CPU Results}

To further test the extent of the above-observed phenomena, we repeated the experiment in the introduction (Section~\ref{sec:introduction}) on 64 cores of an AMD Milan CPU of Perlmutter.
The input size for the matrices on each of the 64 cores is 3344$\times$3344 with 30 repetitions.
The \texttt{gcc} (v11.2) compiler and Cray LibSci library (v21.08) were used.
While similar results were obtained, the extent of variation was much lower than the 67\% observed on the GPU.
The average CPU power consumption with random inputs ($\approx$188 W) is about 30W (19\%) higher than the \emph{fixed input} case ($\approx$158 W), but still only about 67\% of the TDP.
The memory power consumption and the execution time remain almost unchanged.
The aggregate CPU performance is $\approx$2.0 TFLOP/s in both cases.
We can thus infer the AMD EPYC CPU requires less than 15 pJ/FLOP using vectorized FMA instructions.
\begin{figure}[ht]
\centering
\resizebox{0.95\columnwidth}{!}{%
\begin{tikzpicture}
    \begin{axis}[
            ybar,
            bar width=0.8cm,%<- changed
            % width=0.6\textwidth,
            % height=.4\textwidth,
            width=12cm,
            height=6cm,
            legend style={at={(0.5,1)},
            anchor=north,legend columns=-1},
            symbolic x coords={Time {[s]},CPU Power {[W]},Memory Power {[W]}},
            xtick=data,
            nodes near coords,
            nodes near coords align={vertical},
            ymax=250,
            xticklabel style={align=center},
            ylabel={},
            ymajorgrids=true,
            grid style=dashed,
            nodes near coords={\pgfmathprintnumber[zerofill, precision=1]{\pgfplotspointmeta}},
            enlarge x limits={abs=2*\pgfplotbarwidth}
        ]
        \addplot coordinates {(Time {[s]},72.9) (CPU Power {[W]},157.7) (Memory Power {[W]},114.6)};
        \addplot coordinates {(Time {[s]},73.0) (CPU Power {[W]},188.4) (Memory Power {[W]},114.9)};
        \legend{Fixed Input,Random Input}
    \end{axis}
\end{tikzpicture}
}
\caption{Execution time and power for 30 DGEMMs with fixed and random inputs on 64 cores of AMD EPYC CPU (Input Size: 3344$\times$3344 on each core)}
\label{fig:DGEMM_FixedVsRandom_CPU}
\end{figure}
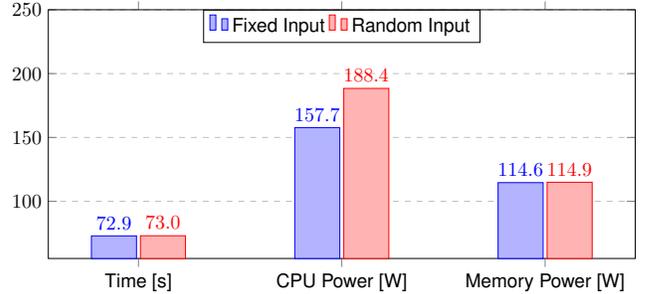

\section{Related Work}
\label{sec:Related}
Measurement and mitigation of variability both in power and performance are important areas of HPC research. 
Power and performance modeling of HPC systems have been explored by several previous works~\cite{tiwari2012modeling,witkowski2013practical,chetsa2014exploiting,wu2016using}.
In the absence of reliable methods to measure power in previous generation architectures, several external meter solutions have been proposed~\cite{bedard2010powermon,dolz2015ardupower,rashti2015wattprof}.
In this work, we have used the hardware interfaces provided by NVIDIA and Cray to obtain high-resolution power measurements.

Performance improvement over several generations of architecture has come at a cost of increased variations in both performance and power.
Variations in performance have been attributed to contention for resources between applications and operating system~\cite{skinner2005understanding}, system utilization~\cite{li2019effect}, network~\cite{bhatele2020case} and I/O~\cite{yildiz2016root} among others.
The run-to-run and variation across processors have been shown to be up to 15\% and 20\% respectively mainly due to manufacturing variability leading to differences in operating frequencies~\cite{marathe2017empirical}. 
Variations in power and performance of GPUs due to hardware or system software have been reported~\cite{debardeleben2013gpu}.
However, none of the above addresses the variation emanating from the program inputs themselves.

Several works have attempted to mitigate the effects of manufacturing variation~\cite{gholkar2015power, inadomi2015analyzing}.
Exploiting the variations in processor and GPU power due to application characteristics to improve energy efficiency have been well studied~\cite{ dblp:conf/ics/rountreelssfb09,wang2015evaluating,bhalachandra2015using,marathe2015run,bhalachandra2017adaptive,adhinarayanan2018making,kumar2021cuttlefish}.
Variations in the node power draw predominantly in non-GPU systems have also been used to an advantage for improving performance~\cite{patki2013exploring,sarood2013optimizing,sarood2014maximizing,sakamoto2017production}.
There have also been attempts to provide end-to-end HPC power management solutions~\cite{eastep2017global,wu2020toward}.
Again, these works do not focus on variations arising due to the actual program inputs explicitly.

The current work shows that program inputs can have a large impact on power and marginal effects on performance, which future policies for maximizing performance within a budget or improving energy efficiency may need to take into consideration.

\section{Discussion \& Conclusion}
\label{sec:Conclusion}
Observations of input-induced variations in power motivated us to investigate how the data sent to a time/thread-multiplexed GPU core strides through memory.
This is different from previous research that investigated how a thread or thread block strides through memory.
The dynamic simultaneous multi-threading (SMT) nature of a streaming multiprocessor makes it hard to analyze analytically.
Thus, we constructed a series of experiments to highlight how changes in input values give rise to a highly variable power.  
GPU DGEMM power is shown to be highly dependent on the entropy of inputs to the FPUs
ranging from roughly 60-100\% of TDP.
This allowed us to provide an upper bound on GPU and CPU FPU energies of 8 and 15 pJ/FLOP respectively --- the difference possibly attributable to the former's use of tensor cores.  

Perhaps one of the more worrying problems is potential inter-process correlations in power.
One can imagine matrix or vector values initially set to zero tending towards effectively random numbers.  This suggests all processes in a job start at lower power and eventually run at high power.  If all the processes of a job are co-located in a rack, rack-to-rack power can swing substantially.  Conversely, if all the processes of a job were spread through a system, rack-to-rack power fluctuations would be mitigated but network bandwidth tapering could become a performance impediment.

The result that power may not always be commensurate with performance is important.
As many moderately sparse systems are treated as dense (zero fill) or systems with duplicate coefficients are treated as if they were unique, our methodology provides insights into the degree to which these applications may be overprovisioned in terms of power.  Avoiding zero fill and exploiting sparsity will reduce run time, but may increase power as only non-zeros are operated on.

The insights from this study may also have implications on performance benchmarking and future power-aware HPC research, especially for architectures that can ramp frequency in response to observed power rather than observed utilization or the floating point (FP) rate.
As roughly 40\% of NVIDIA GPU power was found to be in the datapath (memory through the tensor core), we find there is less than a potential 1.7$\times$ win for iso-power, iso-area, iso-process architectural improvements in NVIDIA GPUs.  Future investigation into 1) how DFMA (double-precision vector FMA) energy per FLOP differs from DMMA (double-precision tensor core multiplication) energy per FLOP and 2) how much energy is consumed in the memory subsystem would provide a tighter bound.  One could certainly imagine extending our benchmarking methodology to capturing energy in the memory, cache, or register files.

Ultimately, we have demonstrated the imperative of including value entropy in power benchmarking to set realistic bounds on node, rack, and system power. 
Moreover, we have utterly dispelled the belief that performance is necessarily correlated with sustained power.
Moving forward, it is imperative that vendors, system integrators, and compute centers adopt benchmarking methodologies that integrate input entropy.

\ifCLASSOPTIONcompsoc
  \section*{Acknowledgments}
\else
  \section*{Acknowledgment}
\fi

This research used resources of the National Energy Research Scientific Computing Center (NERSC), a U.S. Department of Energy Office of Science User Facility located at Lawrence Berkeley National Laboratory, operated under Contract No. DE-AC02-05CH11231.

\section*{Disclaimer}
This report was prepared as an account of work sponsored by an agency of the United States Government. Neither the United States Government nor any agency thereof, nor any of their employees, makes any warranty, express or implied, or assumes any legal liability or responsibility for the accuracy, completeness, or usefulness of any information, apparatus, product, or process disclosed, or represents that its use would not infringe privately owned rights. Reference herein to any specific commercial product, process, or service by trade name, trademark, manufacturer, or otherwise does not necessarily constitute or imply its endorsement, recommendation, or favoring by the United States Government or any agency thereof. The views and opinions of authors expressed herein do not necessarily state or reflect those of the United States Government or any agency thereof.

\ifCLASSOPTIONcaptionsoff
  \newpage
\fi

\bibliographystyle{IEEEtran}
\bibliography{IEEEabrv,bib/main}

\begin{thebibliography}{10}
\providecommand{\url}[1]{#1}
\csname url@samestyle\endcsname
\providecommand{\newblock}{\relax}
\providecommand{\bibinfo}[2]{#2}
\providecommand{\BIBentrySTDinterwordspacing}{\spaceskip=0pt\relax}
\providecommand{\BIBentryALTinterwordstretchfactor}{4}
\providecommand{\BIBentryALTinterwordspacing}{\spaceskip=\fontdimen2\font plus
\BIBentryALTinterwordstretchfactor\fontdimen3\font minus
  \fontdimen4\font\relax}
\providecommand{\BIBforeignlanguage}[2]{{%
\expandafter\ifx\csname l@#1\endcsname\relax
\typeout{** WARNING: IEEEtran.bst: No hyphenation pattern has been}%
\typeout{** loaded for the language `#1'. Using the pattern for}%
\typeout{** the default language instead.}%
\else
\language=\csname l@#1\endcsname
\fi
#2}}
\providecommand{\BIBdecl}{\relax}
\BIBdecl

\bibitem{frontier}
``Top500: Frontier,'' https://www.top500.org/system/180047/, 2022.

\bibitem{kogge2008exascale}
P.~Kogge, K.~Bergman, S.~Borkar, D.~Campbell, W.~Carson, W.~Dally, M.~Denneau,
  P.~Franzon, W.~Harrod, K.~Hill \emph{et~al.}, ``Exascale computing study:
  Technology challenges in achieving exascale systems,'' 2008.

\bibitem{bhalachandra2021understanding}
S.~Bhalachandra, B.~Austin, and N.~J. Wright, ``Understanding power variation
  and its implications on performance optimization on the cori supercomputer,''
  in \emph{2021 International Workshop on Performance Modeling, Benchmarking
  and Simulation of High Performance Computer Systems (PMBS)}.\hskip 1em plus
  0.5em minus 0.4em\relax IEEE, 2021, pp. 51--62.

\bibitem{dblp:conf/ipps/rountreeasls12}
B.~Rountree, D.~H. Ahn, B.~R. de~Supinski, D.~K. Lowenthal, and M.~Schulz,
  ``Beyond dvfs: A first look at performance under a hardware-enforced power
  bound,'' in \emph{26th IEEE International Parallel and Distributed Processing
  Symposium Workshops {\&} PhD Forum, IPDPS 2012, Shanghai, China, May 21-25,
  2012}, 2012.

\bibitem{marathe2017towards}
A.~Marathe, G.~Abdulla, B.~L. Rountree, and K.~Shoga, ``Towards a unified
  monitoring framework for power, performance and thermal metrics: A case study
  on the evaluation of hpc cooling systems,'' in \emph{2017 IEEE International
  Parallel and Distributed Processing Symposium Workshops (IPDPSW)}.\hskip 1em
  plus 0.5em minus 0.4em\relax IEEE, 2017, pp. 974--983.

\bibitem{porterfield2013power}
A.~K. Porterfield, S.~L. Olivier, S.~Bhalachandra, and J.~F. Prins, ``Power
  measurement and concurrency throttling for energy reduction in openmp
  programs,'' in \emph{Parallel and Distributed Processing Symposium Workshops
  \& PhD Forum (IPDPSW), 2013 IEEE 27th International}.\hskip 1em plus 0.5em
  minus 0.4em\relax IEEE, 2013, pp. 884--891.

\bibitem{porterfield2016variability}
A.~Porterfield, S.~Bhalachandra, W.~Wang, and R.~Fowler, ``Variability: A
  tuning headache,'' in \emph{2016 IEEE International Parallel and Distributed
  Processing Symposium Workshops (IPDPSW)}.\hskip 1em plus 0.5em minus
  0.4em\relax IEEE, 2016, pp. 1069--1072.

\bibitem{shalf2020future}
J.~Shalf, ``The future of computing beyond moore’s law,'' \emph{Philosophical
  Transactions of the Royal Society A}, vol. 378, no. 2166, p. 20190061, 2020.

\bibitem{martin2014cray}
S.~J. Martin and M.~Kappel, ``Cray xc30 power monitoring and management,'' in
  \emph{Cray User Group Conference Proceedings}, 2014.

\bibitem{nersc-proxies}
``Nersc {Proxies},'' https://gitlab.com/NERSC/nersc-proxies/info.

\bibitem{tiwari2012modeling}
A.~Tiwari, M.~A. Laurenzano, L.~Carrington, and A.~Snavely, ``Modeling power
  and energy usage of hpc kernels,'' in \emph{2012 IEEE 26th International
  Parallel and Distributed Processing Symposium Workshops \& PhD Forum}.\hskip
  1em plus 0.5em minus 0.4em\relax IEEE, 2012, pp. 990--998.

\bibitem{witkowski2013practical}
M.~Witkowski, A.~Oleksiak, T.~Piontek, and J.~Weglarz, ``Practical power
  consumption estimation for real life hpc applications,'' \emph{Future
  Generation Computer Systems}, vol.~29, no.~1, pp. 208--217, 2013.

\bibitem{chetsa2014exploiting}
G.~T. Chetsa, L.~Lef{\`e}vre, J.-M. Pierson, P.~Stolf, and G.~Da~Costa,
  ``Exploiting performance counters to predict and improve energy performance
  of hpc systems,'' \emph{Future Generation Computer Systems}, vol.~36, pp.
  287--298, 2014.

\bibitem{wu2016using}
X.~Wu, V.~Taylor, J.~Cook, and P.~J. Mucci, ``Using performance-power modeling
  to improve energy efficiency of hpc applications,'' \emph{Computer}, vol.~49,
  no.~10, pp. 20--29, 2016.

\bibitem{bedard2010powermon}
D.~Bedard, M.~Y. Lim, R.~Fowler, and A.~Porterfield, ``Powermon: Fine-grained
  and integrated power monitoring for commodity computer systems,'' in
  \emph{Proceedings of the IEEE SoutheastCon 2010 (SoutheastCon)}.\hskip 1em
  plus 0.5em minus 0.4em\relax IEEE, 2010, pp. 479--484.

\bibitem{dolz2015ardupower}
M.~F. Dolz, M.~R. Heidari, M.~Kuhn, T.~Ludwig, and G.~Fabregat, ``Ardupower: A
  low-cost wattmeter to improve energy efficiency of hpc applications,'' in
  \emph{2015 Sixth International Green and Sustainable Computing Conference
  (IGSC)}.\hskip 1em plus 0.5em minus 0.4em\relax IEEE, 2015, pp. 1--8.

\bibitem{rashti2015wattprof}
M.~Rashti, G.~Sabin, D.~Vansickle, and B.~Norris, ``Wattprof: A flexible
  platform for fine-grained hpc power profiling,'' in \emph{2015 IEEE
  International Conference on Cluster Computing}.\hskip 1em plus 0.5em minus
  0.4em\relax IEEE, 2015, pp. 698--705.

\bibitem{skinner2005understanding}
D.~Skinner and W.~Kramer, ``Understanding the causes of performance variability
  in hpc workloads,'' in \emph{IEEE International. 2005 Proceedings of the IEEE
  Workload Characterization Symposium, 2005.}\hskip 1em plus 0.5em minus
  0.4em\relax IEEE, 2005, pp. 137--149.

\bibitem{li2019effect}
B.~Li, S.~Chunduri, K.~Harms, Y.~Fan, and Z.~Lan, ``The effect of system
  utilization on application performance variability,'' in \emph{Proceedings of
  the 9th International Workshop on Runtime and Operating Systems for
  Supercomputers}, 2019, pp. 11--18.

\bibitem{bhatele2020case}
A.~Bhatele, J.~J. Thiagarajan, T.~Groves, R.~Anirudh, S.~A. Smith, B.~Cook, and
  D.~K. Lowenthal, ``The case of performance variability on dragonfly-based
  systems,'' in \emph{2020 IEEE International Parallel and Distributed
  Processing Symposium (IPDPS)}.\hskip 1em plus 0.5em minus 0.4em\relax IEEE,
  2020, pp. 896--905.

\bibitem{yildiz2016root}
O.~Yildiz, M.~Dorier, S.~Ibrahim, R.~Ross, and G.~Antoniu, ``On the root causes
  of cross-application i/o interference in hpc storage systems,'' in \emph{2016
  IEEE International Parallel and Distributed Processing Symposium
  (IPDPS)}.\hskip 1em plus 0.5em minus 0.4em\relax IEEE, 2016, pp. 750--759.

\bibitem{marathe2017empirical}
A.~Marathe, Y.~Zhang, G.~Blanks, N.~Kumbhare, G.~Abdulla, and B.~Rountree, ``An
  empirical survey of performance and energy efficiency variation on intel
  processors,'' in \emph{Proceedings of the 5th International Workshop on
  Energy Efficient Supercomputing}, 2017, pp. 1--8.

\bibitem{debardeleben2013gpu}
N.~DeBardeleben, S.~Blanchard, L.~Monroe, P.~Romero, D.~Grunau, C.~Idler, and
  C.~Wright, ``Gpu behavior on a large hpc cluster,'' in \emph{European
  Conference on Parallel Processing}.\hskip 1em plus 0.5em minus 0.4em\relax
  Springer, 2013, pp. 680--689.

\bibitem{gholkar2015power}
N.~Gholkar, F.~Mueller, and B.~Rountree, ``Power tuning for hpc jobs under
  manufacturing variations,'' Tech. Rep.

\bibitem{inadomi2015analyzing}
Y.~Inadomi, T.~Patki, K.~Inoue, M.~Aoyagi, B.~Rountree, M.~Schulz,
  D.~Lowenthal, Y.~Wada, K.~Fukazawa, M.~Ueda \emph{et~al.}, ``Analyzing and
  mitigating the impact of manufacturing variability in power-constrained
  supercomputing,'' in \emph{SC'15: Proceedings of the International Conference
  for High Performance Computing, Networking, Storage and Analysis}.\hskip 1em
  plus 0.5em minus 0.4em\relax IEEE, 2015, pp. 1--12.

\bibitem{dblp:conf/ics/rountreelssfb09}
B.~Rountree, D.~K. Lowenthal, B.~R. de~Supinski, M.~Schulz, V.~W. Freeh, and
  T.~K. Bletsch, ``Adagio: {Making DVS} practical for complex {HPC}
  applications,'' in \emph{ICS '09: Proc. of the 23rd Intl. Conference on
  Supercomputing}, 2009.

\bibitem{wang2015evaluating}
B.~Wang, D.~Schmidl, and M.~S. M{\"u}ller, ``{Evaluating the Energy Consumption
  of OpenMP Applications on Haswell Processors},'' in \emph{OpenMP:
  Heterogenous Execution and Data Movements}.\hskip 1em plus 0.5em minus
  0.4em\relax Springer, 2015, pp. 233--246.

\bibitem{bhalachandra2015using}
S.~Bhalachandra, A.~Porterfield, and J.~F. Prins, ``Using dynamic duty cycle
  modulation to improve energy efficiency in high performance computing,'' in
  \emph{Parallel and Distributed Processing Symposium Workshop (IPDPSW), 2015
  IEEE International}.\hskip 1em plus 0.5em minus 0.4em\relax IEEE, 2015, pp.
  911--918.

\bibitem{marathe2015run}
A.~Marathe, P.~E. Bailey, D.~K. Lowenthal, B.~Rountree, M.~Schulz, and B.~R.
  de~Supinski, ``A run-time system for power-constrained hpc applications,'' in
  \emph{International Conference on High Performance Computing}.\hskip 1em plus
  0.5em minus 0.4em\relax Springer, 2015, pp. 394--408.

\bibitem{bhalachandra2017adaptive}
S.~Bhalachandra, A.~Porterfield, S.~L. Olivier, and J.~F. Prins, ``An adaptive
  core-specific runtime for energy efficiency,'' in \emph{Parallel and
  Distributed Processing Symposium (IPDPS), 2017 IEEE International}.\hskip 1em
  plus 0.5em minus 0.4em\relax IEEE, 2017, pp. 947--956.

\bibitem{adhinarayanan2018making}
V.~Adhinarayanan, B.~Dutta, and W.-c. Feng, ``Making a case for green
  high-performance visualization via embedded graphics processors,'' in
  \emph{2018 IEEE International Parallel and Distributed Processing Symposium
  Workshops (IPDPSW)}.\hskip 1em plus 0.5em minus 0.4em\relax IEEE, 2018, pp.
  721--724.

\bibitem{kumar2021cuttlefish}
S.~Kumar, A.~Gupta, V.~Kumar, and S.~Bhalachandra, ``Cuttlefish: library for
  achieving energy efficiency in multicore parallel programs,'' in
  \emph{Proceedings of the International Conference for High Performance
  Computing, Networking, Storage and Analysis}, 2021, pp. 1--14.

\bibitem{patki2013exploring}
T.~Patki, D.~K. Lowenthal, B.~Rountree, M.~Schulz, and B.~R. de~Supinski,
  ``Exploring hardware overprovisioning in power-constrained, high performance
  computing,'' in \emph{Proceedings of the 27th international ACM conference on
  International conference on supercomputing}.\hskip 1em plus 0.5em minus
  0.4em\relax ACM, 2013, pp. 173--182.

\bibitem{sarood2013optimizing}
O.~Sarood, A.~Langer, L.~Kal{\'e}, B.~Rountree, and B.~De~Supinski,
  ``Optimizing power allocation to cpu and memory subsystems in overprovisioned
  hpc systems,'' in \emph{Cluster Computing (CLUSTER), 2013 IEEE International
  Conference on}.\hskip 1em plus 0.5em minus 0.4em\relax IEEE, 2013, pp. 1--8.

\bibitem{sarood2014maximizing}
O.~Sarood, A.~Langer, A.~Gupta, and L.~Kale, ``Maximizing throughput of
  overprovisioned hpc data centers under a strict power budget,'' in
  \emph{Proceedings of the International Conference for High Performance
  Computing, Networking, Storage and Analysis}.\hskip 1em plus 0.5em minus
  0.4em\relax IEEE Press, 2014, pp. 807--818.

\bibitem{sakamoto2017production}
R.~Sakamoto, T.~Cao, M.~Kondo, K.~Inoue, M.~Ueda, T.~Patki, D.~Ellsworth,
  B.~Rountree, and M.~Schulz, ``Production hardware overprovisioning:
  Real-world performance optimization using an extensible power-aware resource
  management framework,'' in \emph{2017 IEEE International Parallel and
  Distributed Processing Symposium (IPDPS)}.\hskip 1em plus 0.5em minus
  0.4em\relax IEEE, 2017, pp. 957--966.

\bibitem{eastep2017global}
J.~Eastep, S.~Sylvester, C.~Cantalupo, B.~Geltz, F.~Ardanaz, A.~Al-Rawi,
  K.~Livingston, F.~Keceli, M.~Maiterth, and S.~Jana, ``Global extensible open
  power manager: a vehicle for hpc community collaboration on co-designed
  energy management solutions,'' in \emph{International Supercomputing
  Conference}.\hskip 1em plus 0.5em minus 0.4em\relax Springer, 2017, pp.
  394--412.

\bibitem{wu2020toward}
X.~Wu, A.~Marathe, S.~Jana, O.~Vysocky, J.~John, A.~Bartolini, L.~Riha,
  M.~Gerndt, V.~Taylor, and S.~Bhalachandra, ``Toward an end-to-end auto-tuning
  framework in hpc powerstack,'' in \emph{2020 IEEE International Conference on
  Cluster Computing (CLUSTER)}.\hskip 1em plus 0.5em minus 0.4em\relax IEEE,
  2020, pp. 473--483.

\end{thebibliography}

\end{document}